\documentstyle[11pt]{article}
\renewcommand{\baselinestretch}{1.25} 
\newfont{\larom}{cmbx10 scaled\magstep3}
\newfont{\bsan}{cmssbx10}
\begin{document}
\begin{center}
  {\larom Observations in the Einstein--de Sitter Cosmology:
	  Dust Statistics and Limits of Apparent Homogeneity}  

  \vspace{10mm}
  {\Large Marcelo B. Ribeiro\footnote{{\bf Current address:} Instituto de
  F\'{\i}sica, Universidade Federal do Rio de Janeiro-UFRJ, C.P.\ 68528,
  Ilha do Fund\~{a}o, Rio de Janeiro, RJ 21945-970, Brazil; E-mail:
  {\tt mbr@if.ufrj.br}.}}\\
  \vspace{5mm}
  {\normalsize Observat\'{o}rio Nacional, Rio de Janeiro, Brazil} 

  \vspace{6mm}
  {\bsan Published in The Astrophysical Journal, 441, 477-487, (1995)}

  \vspace{8mm}
  {\bf ABSTRACT}
\end{center}
  \begin{quotation}
    \small

    The two-point correlation function for the dust distribution in the
    unperturbed Einstein-de Sitter cosmological model is studied along the
    past light cone. It was found that the two-point correlation function
    seems unable to represent the theoretical distribution of dust along
    the backward null cone of this unperturbed model, which has already
    been determined in a previous paper as being apparently inhomogeneous
    at ranges usually considered as local. Such result was revisited in
    order to determine more precisely the quantitative limits where, in
    theory, we can detect apparent homogeneity, and it was found that this
    may only happen up to $z \sim 10^{-2}$. A different statistical
    analysis proposed by Pietronero is used, and it appears to be able to
    represent more accurately the theoretical distribution of dust in this
    cosmology. In the light of these results, it is argued that the usual
    practice of disregarding relativistic effects in studies of distribution
    of galaxies, by considering them as being placed on local regions, seems
    to be valid only on much closer scales than it is commonly believed,
    if the Einstein-de Sitter model is used as a theoretical framework
    for studying such distributions. In this cosmology with
    $H_0=75 \ \mbox{Km} \ {\mbox{s}}^{-1} {\mbox{Mpc}}^{-1}$, that may
    only happen in redshifts as low as $z \approx 0.04$, which means that
    the local approximation seems to be valid up to zeroth order of
    approximation only. As at present there are many redshift surveys
    which have already probed at deeper ranges, it seems that in order
    to compare the Friedmann models with observations we have to be very
    careful when ignoring the past light cone problem in observational
    cosmology, either in theoretical calculations or in data analysis,
    due to relativistic effects which produce observable inhomogeneity
    even in spatially homogeneous cosmological models. 

    \vspace{3mm}


  \end{quotation}
\newpage
\section{Introduction}

The spatial two-point correlation function is perhaps the most popular
statistical tool currently used for the characterization of the large-scale
distribution of galaxies. It is extensively used to extract basic
statistical information from galaxy catalogues, measuring their clustering
properties, information which is then often used by theoreticians in their
cosmological models. The most important result seems to be the finding that
at small scales the spatial two-point correlation function is observed
to follow a power law with negative slope (Peebles 1980), and as the
correlation function decreases towards one or zero when the
distance increases, the distance where it reaches these values is usually
associated with the homogeneization of the sample in the sense that we
are supposed to have reached a homogeneous ``fair sample'' of the
universe (Peebles 1980).

Nevertheless, doubts about the reliability of the correlation function
have been voiced by some authors. Einasto, Klypin \& Saar (1986) and Davis
et al. (1988) found out that the correlation length $r_0$, which is supposed
to indicate the depth where we have reached a ``fair sample'' of the
distribution of galaxies, seems to increase with the sample size (see also
Mart\'{\i}nez et al.\ 1993).
Pietronero (1987) criticized the use of the correlation function
in the context of the distribution of the galaxies under the grounds that
it is conceptually incorrect since it could only be applied to physical
systems whose average densities are well defined quantities, property,
his argument goes, apparently not found up to now in the observed
distribution of galaxies (see also Coleman, Pietronero \& Sanders 1988). Geller
(1989) voiced a similar concern as her slices have inhomogeneities that are
large compared with the sample volume. In addition, Coleman \& Pietronero
(1992) carried out simulations of some computer generated samples and
found out that the spatial two-point correlation function seems to be
unable to describe correctly the samples studied, whose distribution
were known by construction. Such result is claimed by them to give
support to the hypothesis that the observed large-scale distribution of
galaxies does seem to follow a fractal pattern. Finally, by using a
different from usual correlation analysis, Coleman, Pietronero \& Sanders
(1988) reported that there is no indication so far of any correlation
length in the CfA samples, that is, no indication of a homogeneous
``fair sample'' being reached, which seems to contradict the results
obtained from the measurements of the usual correlation function where a
homogeneization of the distribution of galaxies is indicated.

Another important aspect that remains basically untested is the
question of at what scales the correlation function itself may be
affected by the curvature and expansion of the universe, and the
answer to this question might impose important limits in its usefulness.
In particular, it is necessary to determine quantitatively the limits
of observational detectability of the spatial homogeneity of the
popular spatially homogeneous Friedmann models, since observations are
carried out along our past light cone and what we can directly observe are
galaxies placed at different distances at different times. This last
remark is of special importance since Ribeiro (1992) showed that if the
universe is really Einstein-de Sitter, departures from local apparent
homogeneity should be observed at much closer scales than it is usually
assumed.

This paper attempts to address the issues surrounding this controversy
from a different perspective than the authors mentioned above. By
starting from the most popular relativistic cosmological model, that is,
the Einstein-de Sitter model, I study its unperturbed behaviour along
the past light cone and obtain some theoretical predictions which can be
used to analise results obtained through actual astronomical observations.
Therefore, I do not make the usual assumption that relativistic
effects can be ignored and Euclidean geometry can be used when dealing
with the typical depths of the current redshift surveys. On the contrary,
I intend to ascertain to what extent this hypothesis may be really a
good approximation.

The aim of this paper is twofold. First to further analise the basic
result of Ribeiro (1992), which, simply speaking, states that relativistic
corrections do matter in cosmology, even at small scales. In special,
one of the main goals here is to study the relativistic effects on
some statistical tests used in the study of large-scale distribution of
galaxies. Second, to attempt to set up explicit quantitative limits to
the observation of spatial homogeneity in the unperturbed Einstein-de Sitter
model, the ranges of observational detectability of the homogeneous
hypothesis, relating
those limits to the error margins associated with astronomical
measurements, and trying to see which statistical test best reproduces
the actual distribution of dust in this model.

This paper is organized as follows. In \S 2 I present a brief summary of the
basic observational relations necessary for the model under study, which
were obtained in Ribeiro (1992), and in \S 3 I derive the correlation
function and other functions along the past light cone. Section 4 shows an
analysis of the results obtained in the previous section, presents 
some limits to the observation of apparent homogeneity in the
unperturbed Einstein-de Sitter model, and discuss how the
correlation function seems to be an inadequate
tool to characterize the distribution of galaxies, while the statistical
analysis advanced by Pietronero (1987) appears to describe more
accurately the actual distribution of dust in this model. The paper
finishes with a discussion on the various possible implications of
these results.

\section{Observational Relations in the Einstein-de Sitter Spacetime}

Let us write the Einstein-de Sitter metric (with $c=G=1$, $\Lambda =0$) in
the form
\begin{equation}
   dS^2 = dt^2 - a^2(t) \left[ dr^2 - r^2 \left( d \theta^2 + \sin^2
          \theta d \phi^2 \right) \right],
 \label{1}
\end{equation}
where the function $a(t)$ is given by
\begin{equation}
  { \left( \frac{da}{dt} \right) }^2 = \frac{8 \pi}{3} \rho a^2
  \label{2}
\end{equation}
and the local density is
\begin{equation}
  \rho = \frac{1}{6 \pi a^3(t)}.
  \label{3}
\end{equation}
If we label $t=0$ as ``now'', that is, $t=0$ being defined as our
present time hypersurface, the solution of equation (\ref{2}) may be
written as
\begin{equation}
  a(t)= { \left( t + \frac{2}{3 H_0} \right) }^{2/3},
  \label{4}
\end{equation}
where $H_0$ is the present value of the Hubble constant.

We are interested in studying this metric along the past light cone
inasmuch as this hypersurface is where astronomical observations are
actually made, and for this purpose it becomes necessary to integrate
the past radial null geodesic of metric (\ref{1}),
\[
   \frac{dt}{dr} = -a(t),
\]
from ``here and now'' ($t=r=0$) till $t(r)$. The result may be
written as
\begin{equation}
  3 { \left( t + \frac{2}{3 H_0} \right) }^{1/3} = 
  { \left( \frac{18}{H_0} \right) }^{1/3} -r.
  \label{5}
\end{equation}
Here I have chosen to use the radius coordinate $r$ as the parameter
along the null geodesic.

The observational relations necessary in this paper have
already been calculated in Ribeiro (1992). Along the backward null cone
the cumulative number count $N_c$, the luminosity distance $d_\ell$, the
redshift $z$, the observed volume $V$ and the average density $\langle
\rho \rangle$ are respectively given by
\begin{equation}
   N_c =  \frac{2r^3}{9 M_G},
   \label{6}
\end{equation}
\begin{equation}
   d_\ell =  9r { \left( \frac{2}{3 H_0} \right) }^{4/3}
	   { \left[ { \left( \frac{18}{H_0} \right) }^{1/3} -r \right]
	   }^{-2},
   \label{7}
\end{equation}
\begin{equation}
   1+z={ \left( \frac{18}{H_0} \right) }^{2/3}  { \left[ { \left(
   \frac{18}{H_0} \right) }^{1/3} -r \right] }^{-2}, 
   \label{8}
\end{equation}
\begin{equation}
   V \equiv \frac{4}{3} \pi {d_\ell}^3 = \frac{192 \pi r^3}{{H_0}^4 { \left[
   { \left( 18/H_0 \right) }^{1/3} - r \right] }^6},
   \label{9}
\end{equation}
\begin{equation}
   \langle \rho \rangle  \equiv \frac{N_c M_G}{V} = \frac{{H_0}^4}{864 \pi} {
   \left[ { \left( \frac{18}{H_0} \right) }^{1/3} -r \right]
	      }^6,
   \label{10}
\end{equation}
where $M_G$ is the average galactic rest mass ($\sim 10^{11}
M_{\odot}$). 

If we use the inverse of equation (\ref{7}), the cumulative number
counting and the average density may be written
in terms of the luminosity distance as
\begin{equation}
   N_c =  \frac{4}{H_0 M_G} { \left[ 1- { \left( \frac{1}{2} + \sqrt{
   \frac{H_0 d_\ell}{2} + \frac{1}{4}} \ \right) }^{-1} \right] }^3,
   \label{11}
\end{equation}   
\begin{equation}
   \langle \rho \rangle = \frac{3 {H_0}^2}{8 \pi}
			  { \left( \frac{1}{2} + \sqrt{ \frac{H_0
			  d_\ell}{2} + \frac{1}{4}} \ \right) }^{-6}.
   \label{12}
\end{equation}

Before closing this section, a few words are necessary here in order
to explain why the luminosity distance $d_\ell$ was chosen as the
measurement of distance in this paper. As is well known, in
relativistic cosmology we do not have an unique way of measuring the
distance between source and observer since such measurement depends
on circumstances. We can, for instance, make use of geometrically
defined distances like the proper radius, or observationally defined
distances like the luminosity distance or the observer area distance
(also known as angular diameter distance) in order to say that a
certain object lies at a certain distance from us. The circumstances
which tell us which definition to use can also be determined on
observational grounds, and so if we only have at our disposal the
apparent magnitudes of galaxies we associate to each of them the
luminosity distance and use such measurement in our analysis. On the
other hand, if these apparent magnitudes are corrected by the redshift
of the sources, then we can associate the corrected luminosity distance,
which is the same as the observer area distance obtained if we have the
apparent size of the objects (Ellis 1971), and, therefore, another kind of
distance measure is obtained. Any of these observational distances
is as valid as any other, as real as any other, with the choice
being dictated by the availability of data, the nature
of the problem being treated and its convenience, but they will only
have the same value at $z \ll 1$, varying, sometimes widely, for larger
$z$ (see  McVittie 1974 for a comparison of these distances in simple
cosmological models).

In this paper we are interested in observables because we seek to
compare theory with observations, and this means that geometrical
distances are of no interest here. Consequently, the approach of this
paper is different from others where unobservable coordinate distances
(differences between coordinates) and separations (integration of the
line element $dS$ over some previously defined surface) are taken as
measure of distance, and in order to develop a treatment coherent with
the observational approach of this problem we need to make a choice
among the observational distances based on the nature of the problem and
the observations available.

In this work I intend to study the theoretical distribution of density
and the correlation function obtained in the Einstein-de Sitter model
in order to compare their forms with the ones produced by the recent
all-sky redshift surveys, and in this field it is usual for observers to take 
the luminosity distance as their indicator of distance (for example,
Saunders et al.\ 1990 do use $d_\ell$ in their extensive statistical
analysis of a sample of IRAS galaxies). It seems therefore perfectly
reasonable to take  the luminosity distance as the most appropriate
definition of distance to use in the context of this work, because
what is sought is to mimic the current methodology used by many
observers in this field, and to carry out a comparison between the
theoretical predictions of this model and the observational results
brought by the redshift surveys.

\section{Dust Statistics Along the Past Light Cone}

Once we have established the basic observational relations which
shall be needed here, the next step is to use those relations in order
to find the expression of the correlation function for the distribution
of dust in the Einstein-de Sitter model. Nonetheless, before we go into
the details of the calculations themselves, a few remarks about the
physical behaviour of the model under study will be helpful in order to
make clear the meaning of the results obtained.

The Einstein-de Sitter model is spatially homogeneous, and this is obvious
from equation (\ref{3}) since each specific value of the time coordinate
corresponds to another specific and constant value for the local density.
Therefore, at hypersurfaces of constant time, that is, at spatial sections
of the model, the density remains unchanged, and at our present time
hypersurface the local density is given by the constant
\begin{equation}
   \rho_0  = \frac{3 {H_0}^2}{8 \pi}.
   \label{13}
\end{equation}
It is well known that this result cannot be accepted at its face value 
since astronomical observations indicate lumpiness of matter, and a constant
local density would mean that there would not be galaxies, at least at
smaller scales. As a consequence of this, the current view is to assume that
there are metric perturbations such that local density fluctuations
$\delta \rho / \rho$ arise and produce the observed lumpiness of matter.
Then the current thinking assumes that beyond certain length scale the density
fluctuations will decrease, and eventually approach asymptotically the
value $\rho_0$. Such reasoning naturally leads to a statistical 
interpretation of equation (\ref{13}) in the sense that $\rho_0$ would be the
average value of the local density fluctuations, and, under this view,
the cosmological problem becomes the determination of the value of
$\rho_0$ and the density perturbations scale length. This discussion
is also valid for the other Friedmann models, either open or closed,
with one of the differences being the value of $\rho_0$, but in this
paper I shall only deal with the Einstein-de Sitter model, for reasons of
simplicity and analytical feasibility.

The spatial two-point correlation function is the most common statistical
tool currently used with the aim of studying the cosmological density
fluctuations. It assumes that the objects under discussion (galaxies) can be
regarded as point particles that are distributed homogeneously on
sufficiently large scale. This means that we can meaningfully assign an average
number density to the distribution and, therefore, we can characterize the 
galaxy distribution in terms of the extent of the departures from uniformity
on various scales. Consequently, the correlation function would, in
principle, give us a methodology by which one hopes to achieve both
aims as described above: to give the approximate density perturbation
scale length and, indirectly, the value of $\rho_0$ (not necessarily
the specific one given by the Einstein-de Sitter model, eq.\  [\ref{13}]),
and that would be indicated at the ranges where the correlation function
would tend to zero, meaning the approach to a Poisson distribution and,
therefore, the expected homogeneity at deeper scales. Once the range of
homogeneity is established we would have achieved a ``fair sample'' of
the Universe and we could, in theory, calculate $\rho_0$ by means of,
say, counts of galaxies beyond the fair sample range.

Such scheme, however, usually disregards an essential point: observations
are carried out along the past light cone and, hence, relativistic
corrections might become important at larger scales. In particular, inasmuch 
as along the past light cone the proper density changes due to the 
crossing of the null geodesic through hypersurfaces of constant time, but
with different values of the proper density, a cumulative average density
will change for the same reason and, therefore, we will end up having a
not well defined average number density for the distribution of dust even
in Friedmann cosmologies (Ribeiro 1992, 1993). In other words, since the
past light cone is a hypersurface of inhomogeneity (the density varies
along it), as far as observations are concerned, in the spatially
homogeneous Friedmann models their geometrical homogeneity is
in fact a local apparent homogeneity, which becomes an apparent
inhomogeneity for deeper observational ranges. This physical property of
the models is related to the fact that astronomical observations are made
along the special null hypersurface which is in fact inhomogeneous.
There is no contradiction in the fact that a spatially homogeneous model
does have hypersurfaces of inhomogeneity, as it continues to exhibit
the physical property of having constant density at constant time
slices. Therefore, it is important to state clearly that the Einstein-de
Sitter model is not only {\it spatially homogeneous}, but also {\it
apparently inhomogeneous}. This effect of observational inhomogeneity of
the model may also be termed as ``evolution of the background''.

The implicit answer to the points raised above is to argue that at
small scales, that is, at small redshifts, relativistic effects are not
important, and so we can carry out calculations and analysis of data
using relations valid only along our present time hypersurface, and
ignore the null
geodesic problem because the values of $z$ of the all-sky redshift
surveys are supposed to be small, and would produce insignificant
relativistic corrections in the observables. In other words, since at
small scales the apparent inhomogeneity turns out to be local apparent
homogeneity, then, the reasoning goes, the null geodesic could be
ignored. Nevertheless, putting the problem in that way the
key question remains hidden: how small should the scales be such that
we can ignore the backward null cone? In other words, we need a criterion
to tell us {\it quantitatively} to what extent we can still
consider small scales as small such that we could then safely disregards
the past light cone problem up to that range, or still putting the
problem differently, we need to know the scales where we would no
longer be receiving photons from objects located at our own present
time hypersurface (Ellis 1987, p.\ 62; MacCallum 1987, p.\ 135). The
correlation function does not help to answer this question as it
already {\it assumes} that homogeneity will be reached at certain
scale, and so it is unable to test the local homogeneous hypothesis itself
(see Pietronero 1987, and Coleman \& Pietronero 1992 for a full
discussion). This is an essential point because it has already been
indicated in a previous paper (Ribeiro 1992) that such a departure from
our local region may happen at redshifts as low as $z \approx 0.04$,
value which is usually considered as well within our local region, and
in that case the implications for the measurement of the galaxy and
cluster correlation functions may be important.\footnote{ \ By local
region I mean the region along the past light cone where observational
relations evaluated at $t=\mbox{now}$ are still a valid approximation
when data error margins are considered.} Equation (\ref{12}) shows
clearly that the Einstein-de Sitter model does not seem to remain
homogeneous along the past light cone (see Ribeiro 1992 for a more
detailed physical discussion about this behaviour, and Ribeiro 1993 for
its extension to the other Friedmann models), and as said above, once
the average density stops having a well defined value, beyond a certain
length scale, we can no longer assign a well defined average number
density to the distribution of galaxies, which means a break down of
the usual correlation function. Bearing this discussion in mind we can
now proceed to determine the expression of this function for the
distribution of dust in the unperturbed Einstein-de Sitter model.

I shall approach the correlation function $\xi(d_\ell)$ through the
``conditional density'' $\Gamma (d_\ell)$, which is simpler to calculate
in this context. According to Pietronero (1987; see also Coleman \&
Pietronero 1992), $\Gamma (d_\ell )$ is defined by
\begin{equation}
  \Gamma (d_\ell) = \frac{1}{S} \frac{d N_c}{d (d_\ell) },
  \label{14}
\end{equation}
where $ S ( d_\ell ) \equiv 4 \pi { d_\ell }^2 $ is the area of the observed
spherical shell of radius $d_\ell$. This equation shows that the
conditional density actually measures the average density at a distance
$d_\ell$ from an occupied point.\footnote{ \ Actually, some of
Pietronero's (1987) treatment was anticipated by Wertz (1970, p.~43-45),
where the definition of a ``conditional density'' appears under the name
``differential density'' (Wertz 1970, p.~17; see also Wertz 1971).}
Taking into account equation (\ref{11}), the conditional density for the
distribution of dust in the Einstein-de Sitter model yields
\[
 \Gamma (d_\ell) = \left( \frac{3{H_0}^2}{\pi M_G \sqrt{1+2 H_0 d_\ell}} \right)
		   \left[ 
			 \left( 1+ H_0 d_\ell \right)
			 \left(4+8 H_0 d_\ell +{H_0}^2 {d_\ell}^2 \right)
		  + \right.
\]
\begin{equation}
		  { \left. + \left(2+ H_0 d_\ell \right)
			   \left(2+ 3 H_0 d_\ell \right)
		  \sqrt{1+2 H_0 d_\ell} \right] }^{-1}.
  \label{15}
\end{equation}
We can also define a ``conditional average density'' (Pietronero 1987;
Coleman \& Pietronero 1992)
\begin{equation}
  \Gamma^\ast (d_\ell ) = \frac{1}{V} \int_V \Gamma (d_\ell) d V =
   \frac{3}{{d_\ell}^3} \int_0^{d_\ell} x^2 \Gamma (x) dx,
   \label{16}
\end{equation}
which gives the behaviour of the average density of a sphere of radius
$d_\ell$ centered around an occupied point averaged over all occupied
points. By means of equations (\ref{15}) and (\ref{16}) the conditional
average density is found to be
\[
 \Gamma^\ast (d_\ell) = \frac{3}{{d_\ell}^3} \lim_{b \rightarrow 0}
      \left\{ \frac{1}{\pi M_G b^3 {d_\ell}^3 {H_0}^4}
	      \left[ \left( 2+3bH_0 \right) \left( 2+bH_0 \right)
		     {d_\ell}^3 \sqrt{1+2bH_0} +  \right.  \right.
\]
\[
                     + { \left( 2+3H_0 d_\ell \right) }^2 b^3
		     - \sqrt{1+2H_0d_\ell} \left( 2+3H_0 d_\ell \right)
		     \left( 2+ H_0 d_\ell \right) b^3 - 
\]
\[
       \left. \left. -9 b^2 {H_0}^2 {d_\ell}^3 - 12 b H_0 {d_\ell}^3
		     -4 {d_\ell}^3
              \right]
      \right\},
\]
and once this limit is performed we get
\[
 \Gamma^\ast (d_\ell) = \left( \frac{3}{\pi M_G {H_0}^4 {d_\ell}^6} \right)
			\left[
			\left( 1+ H_0 d_\ell \right)
			\left(4+8 H_0 d_\ell +{H_0}^2 {d_\ell}^2 \right)
			- \right.
\]
\begin{equation}
			\left. - \left(2+H_0 d_\ell
			\right) \left( 2+ 3 H_0 d_\ell \right)
			\sqrt{1+2H_0 d_\ell} \ \right].
 \label{gamstar}
\end{equation}
After some algebraic manipulation we find that in the limit when
$d_\ell \rightarrow 0$, $\Gamma^\ast \rightarrow ( 3 {H_0}^2 / 8 \pi M_G )$,
as one expects.

As shown by Pietronero (1987; see also Mart\'{\i}nez et al.\ 1993),
the correlation function $\xi (d_\ell)$ is related to $\Gamma (d_\ell)$
in a simple way:
\begin{equation}
 \xi = \frac{ \Gamma}{ \langle n \rangle } - 1,
 \label{18}
\end{equation}
where $\langle n \rangle \equiv \langle \rho \rangle / M_G$ is the
average number density. For the model under study we have
\begin{equation}
  \langle n \rangle = \frac{3 {H_0}^2}{8 \pi M_G} { \left( \frac{1}{2} +
		      \sqrt{ \frac{H_0 d_\ell}{2} + \frac{1}{4}} \
		      \right) }^{-6}.
  \label{19}
\end{equation}

One can verify that equations (\ref{gamstar}) and (\ref{19}) are equal, 
which comes as no surprise since, by its own definition,
$\Gamma^\ast$ indeed describes the average number density. Therefore, we may
write the identity
\begin{equation}
 \Gamma^\ast = \langle n \rangle.
 \label{average}
\end{equation}

This result is in agreement with the point already made by
Coleman \& Pietronero that ``for a test of general tendencies, like
homogeneity versus power law correlations, this ($\Gamma^\ast$ function)
is by far the best test'' since ``it correctly reproduces global
properties which are the object of the present discussion'' (Coleman \&
Pietronero 1992, p.\ 328). In fact, inasmuch as the average number
density $\langle n \rangle$ gives us the form of the distribution of
dust along the backward null cone in the Einstein-de Sitter model,
equation (\ref{average}) is actually telling us that the conditional
average density $\Gamma^\ast$ is an appropriate function to be used
in galaxy catalogues in order to see if the large-scale distribution
of galaxies does follow a dust pattern as given by $\langle n \rangle$
in the Einstein-de Sitter cosmological model.

It is worth writing the power series expansion of equation (\ref{19}),
\begin{equation}
 \langle n \rangle = \frac{3 {H_0}^2}{8 \pi M_G} \left( 1 - 3 H_0 d_\ell
		     + \frac{27}{4} {H_0}^2{d_\ell}^2 - \frac{55}{4}
		     {H_0}^3 {d_\ell}^3 + \ldots \right),
  \label{serie}
\end{equation}
and this expression already shows very clearly that the overall number
density is only constant at very small luminosity distances, that is, at
the zeroth order of approximation. In addition, its zeroth order term is
equal to equation (\ref{13}), apart from a constant, as it should be.

In order to use equation (\ref{19}) to evaluate the correlation function, we
have to suppose that there are $m$ objects confined to a fixed region
$R$ of volume $V_u$ (Peebles 1980, p.\ 145). Hence, the expression for
the two-point spatial correlation function for the dust distribution in
the Einstein-de Sitter model, and along the past light cone, is given by
\[
  \xi( d_\ell ) = \left[ \left( R^3 {H_0}^2+9R^2H_0+12R-6{H_0}^2{d_\ell}^3-
		  19H_0{d_\ell}^2 -16d_\ell \right) H_0 \sqrt{1+2 H_0
		  d_\ell} \ + \right.
\]
\[
		  +\left( 2+RH_0 \right) \left(2+3RH_0 \right)
		  \sqrt{\left( 1+2H_0d_\ell \right) \left( 1+2RH_0
		  \right) }- \left( 1+H_0d_\ell \right)
		  \left( 1+2H_0d_\ell \right) \times
\]
\[
		  \left. \times \left( 4+8H_0d_\ell + {H_0}^2 {d_\ell}^2
		  \right) \right]
		  \left\{ \left( 1+2 H_0 d_\ell \right) \left[
		  \left(2+H_0 d_\ell \right) 
		  \left(2+3 H_0 d_\ell \right) \times \right. \right.
\]
\begin{equation}
                  \times { \left. \left.
		  \sqrt{1+2 H_0 d_\ell} + 
		  \left(1+H_0d_\ell \right)
		  \left(4+8H_0d_\ell+{H_0}^2{d_\ell}^2 \right)
		  \right] \right\} }^{-1},
		  \label{20}
\end{equation}
where the dependence on the sample size $R$ is explicit.

As a final remark before the end of this section, it is important to
stress the explicit dependence of all functions with the luminosity
distance, an effect which comes from the fact that along the past light
cone the spatially homogeneous Einstein-de Sitter model is apparently
inhomogeneous, because the proper density changes since equation
(\ref{3}) is a function of time (Ribeiro 1992). Therefore, it is
essential to determine quantitatively the length scale where the
apparent inhomogeneities start to play a significant role in the
observational quantities above.

\section{Analysis}

As discussed in the previous section, it is usually assumed that at
small redshifts the backward null cone can be ignored and Euclidean
geometry may be used as a good approximation in the study of
distribution of galaxies (see e.g.\ Peebles 1980, p.\ 143). Such
reasoning seems perfectly reasonable once the linear redshift-distance
law represents well the observations at small redshifts. Indeed, the
redshift-distance relation along the past light cone in the unperturbed
Einstein-de Sitter model is easily obtainable from equations
(\ref{7}) and (\ref{8}) as being
\begin{equation}
 d_\ell = \frac{2}{H_0} \left( 1 + z - \sqrt{1+z} \right),
 \label{21}
\end{equation}
whose power series expansion reads
\begin{equation}
 d_\ell = \frac{z}{H_0} + \frac{z^2}{4 H_0} - \frac{z^3}{8 H_0} + \ldots
 \label{22}
\end{equation}
Therefore, the Hubble law is clearly the first order
approximation of equation (\ref{21}), and if we take $H_0=75 \ \mbox{Km}
\ {\mbox{s}}^{-1} {\mbox{Mpc}}^{-1}$ it is easy to see that at $z=0.1$
the contribution of the second order term in equation (\ref{22}) is
about 2.4\% of the total in equation (\ref{21}).~\footnote{ \ From
now on I shall assume $H_0=75 \ \mbox{Km} \ {\mbox{s}}^{-1}
{\mbox{Mpc}}^{-1}$ unless stated otherwise. Units are such that $c=G=1$,
and so distance is given in gigaparsecs ($10^9$ pc), mass in units of
$2.09 \times 10^{22} \ M_{\odot}$ and time in units of 3.26 Gyr
($1 \mbox{yr} = 3.16 \times 10^7$ s).} Even at $z=0.5$ the second
order term contributes to only around 11\%, which means the linear
approximation of equation (\ref{21}) can be considered as valid up to at
least $z \approx 0.5$ once error margins of the same magnitude are
considered. This reasoning is perfectly standard and is being repeated
here just for the sake of clarity of the results which will follow.

The surprising aspect of the analysis of the observational relations
along the past null cone is the effect of small redshifts on the
average number density. Considering equations (\ref{19}) and
(\ref{21}) we can write $\langle n \rangle$ as a function of the redshift,
\begin{equation}
 \langle n \rangle = \frac{3 {H_0}^2}{8 \pi M_G} \frac{1}{ { \left( 1+z
                     \right) }^3 }.
 \label{23}
\end{equation}

Figure 1 shows a plot of equation (\ref{23}) and it is clear the overall
density decreases quite sharply as $z$ increases. A 10\% drop in
$\langle n \rangle$, from the value at our present time hypersurface,
occurs at $z=0.036$ ($\approx 145$ Mpc), which is within the limits
of many redshift surveys. At $z=0.07$ ($\approx 285$ Mpc, depth of
IRAS galaxies), the density is around 80\% from its value now ($t=0$),
and at $z=0.1$ the density drops to 75\% from its initial value. It is
therefore clear that even accepting an error margin of 25\% in the
measurements of the global density, a redshift equals to 0.1 is
approximately the deepest scale where we could observe a homogeneous
distribution of dust in an Einstein-de Sitter model ($z=0.1
\Rightarrow d_\ell \approx 410$ Mpc if
$H_0=75 \ \mbox{Km} \ {\mbox{s}}^{-1} \ {\mbox{Mpc}}^{-1}$;
for $H_0=100 \ \mbox{Km} \ {\mbox{s}}^{-1} \ {\mbox{Mpc}}^{-1}$,
$d_\ell \approx 310$ Mpc). Note that at $z=0.1$ 
the redshift-distance law is still well approximated by a linear
function, with an error margin of less than 2.5\%. At $z=0.5$, the
average number density $\langle n \rangle$ goes down to only 30\% from
its initial value at $t=0$. Consequently, by this method, that is, by
using the average density as a gauge, we are able to determine {\it
quantitatively} the scales where the local apparent homogeneity still
holds, and the few numbers above show very clearly that the end of the
observable homogeneous region in the model under study may happen at
scales much smaller than usually assumed, being not only already inside
the limits of many redshift surveys, but also well within the region of
validity of the Hubble law. Table 1 summarizes the numerical estimates
given above.

Let us see now how the correlation function is affected when measured
along the past light cone. The power series expansion of equation
(\ref{20}) is
\begin{equation}
 \xi(d_\ell)= \frac{ \left( \Xi - 4 \right)}{8} - \frac{ \left( \Xi + 4
               \right) }{2} H_0 d_\ell \left[ 1 - \frac{45}{16} H_0
	       d_\ell + \frac{55}{8} {H_0}^2 {d_\ell}^2 -
	       \frac{1001}{64} {H_0}^3 {d_\ell}^3 + \ldots \right],
   \label{26}
\end{equation}
where
\begin{equation}
  \Xi(R) \equiv \sqrt{1+2RH_0} \left( 2+ RH_0 \right) \left( 2+ 3 RH_0
		\right) + R^3 {H_0}^3 + 9 R^2 {H_0}^2 + 12 R H_0.
  \label{27}
\end{equation}
As in the case of the average number density only the zeroth order term
is constant. This is obviously caused by the fact that the spatially
homogeneous Einstein-de Sitter model is only apparently homogeneous, and
at very low observational distances. Note, however, that even this
zeroth order term depends on the size of the sampling, as pointed out
by Pietronero (1987), and has no resemblance to the observed form of the
correlation function at very small distances, known to follow a power-law
form (Peebles 1980). This last remark is hardly surprising since the unperturbed
Einstein-de Sitter model does not model the small-scale lumpiness of
matter, and therefore this unperturbed model cannot reproduce the
observed correlation function. Nonetheless, at larger scales $\xi(d_\ell)$
seems to present some odd effects. Figure~2 shows the correlation
function (\ref{20}) plotted for different values of the sample size
$R$, and it is clear that even at the regions where the model is
apparently homogeneous, as shown in figure~1, $\xi(d_\ell)$ seems unable
to fully characterize it. With $R=50$ Mpc ($z \approx 0.013$),
$\xi (d_\ell)$ seems to indicate a breakdown of the homogeneous pattern
starting around 10~Mpc ($z \approx 0.003$), which clearly
contradicts equation (\ref{23}) and its plot in figure 1, since both
indicate a continuation of the local homogeneous region to at least
145 Mpc ($z \approx 0.036$), with a 10\% error. For different values
of $R$ a similar situation happens, and although we have seen that
equation (\ref{23}) shows an unique and clear scale length where the
observable homogeneity stops, the  correlation function does not seem to
be able to characterize such distance. For bigger values of $R$ the
results obtained from the correlation function become meaningless since
according to equation (\ref{23}) the average number density stops being
a well defined quantity, and therefore, we can no longer make use of the
correlation function in this context. Hence, the application of the
correlation function in this context appears to be severely limited to
only the linear region with zero slope of figure~1, with its limit being
determined by the error margin considered.~\footnote{ \ For example, for
a 10\% error the correlation function can only be used up to
$z \approx 0.04$. For an error around 25\%, up to $z \approx 0.1$, etc.}
It is also interesting to note that the inflexion of $\xi(d_\ell)$ in
figure~1 seems to be related to the value of $R$ chosen in each plot,
and this is consistent with Pietronero's (1987) critique of the
correlation function in the sense that it mixes up the physical
properties of the samples studied with their sizes. Hence, although the
use of the correlation function in this unperturbed model does not
relate the results to actual observations, it seems clear that the
correlation function is a poor statistical test since if the Universe
were {\it exactly} Einstein-de Sitter, the correlation function would have
difficulties to characterize the global behaviour of the dust in the
model.

In contrast to the correlation function, the conditional density
$\Gamma (d_\ell)$ seems to be able to better trace the dust distribution
in the Einstein-de Sitter model. To see this let us first carry out a
power series expansion in equation (\ref{15}). The result may be written as 
\begin{equation}
 \Gamma (d_\ell) = \frac{3 {H_0}^2}{8 \pi M_G} \left( 1 - 4 H_0 d_\ell
		     + \frac{45}{4} {H_0}^2{d_\ell}^2 - \frac{55}{2}
		     {H_0}^3 {d_\ell}^3 + \ldots \right),
  \label{28}
\end{equation}
where one can see that the zeroth order term is the local density at
the present time (\ref{13}), apart from a normalizing constant, and is the
same factor appearing in equation (\ref{serie}). As in the case of the
average number density $\langle n \rangle$, one can see in equation
(\ref{28}) that deviations from local homogeneity are a first order
effect. Figure~3 shows a plot for equation (\ref{15}) where it is clear
the breakdown of the local apparent homogeneity as the luminosity distance
increases, the same effect shown in figure~1.

However,in the sense of results that could be derived directly from
galaxy catalogues, $\Gamma (d_\ell)$ does not seem to be as a good
quantitative tracer for the behaviour of the average density
$\langle n \rangle$ as the conditional average density
$\Gamma^\ast (d_\ell)$ appears to be. This happens because $\Gamma^\ast
(d_\ell)$ traces $\langle n \rangle$ exactly (eq.\ \ref{average}) while
$\Gamma (d_\ell)$ does not, and hence $\Gamma^\ast (d_\ell)$ is
certainly the function to be calculated from galaxy catalogues in order
to probe the behaviour of the mean number density $\langle n \rangle$.
Table 2 presents the percentage drop of $\Gamma (d_\ell)$ and $\langle n
\rangle$ with increasing redshifts when they are compared with their
values at the present time hypersurface $t=0$. Since $\langle n \rangle$
gives the actual distribution of dust of the model, table 2 shows that
although $\Gamma (d_\ell)$ reproduces the general tendency for the drop
in the density, it is not much sensitive to the drop itself. That seems
to happen because, by its own definition, $\Gamma (d_\ell)$ is more
affected by local fluctuations since it measures the average density
at a distance $d_\ell$ from an occupied point, while $\Gamma^\ast
(d_\ell)$ does this but also averages over all occupied points. Clearly
the latter function is not appropriate when one is investigating local
fluctuations because in that case these fluctuations would be smoothed
out in the average. Nevertheless, $\Gamma^\ast$ would be the desired
test to be used when one is interested in global tendencies and
properties, which is the case when one is studying the 
limits of the apparent homogeneity of the Friedmann cosmologies.

Despite those limitations, the conditional density can still be used in
statistical analyses of the local region of the model. To see this, let
us find the ratio between $\Gamma^\ast$ and $\Gamma$. By using equations
(\ref{15}) and (\ref{gamstar}) one can show that
\begin{equation}
 \frac{\Gamma^\ast}{\Gamma} = \sqrt{1+2H_0 d_\ell}.
 \label{29}
\end{equation}
By means of equation (\ref{21}) we may write the equation above as a
function of the redshift,
\begin{equation}
 \frac{\Gamma^\ast}{\Gamma} = 2 \sqrt{1+z} -1,
 \label{30}
\end{equation}
whose power series expansion yields,
\begin{equation}
 \frac{\Gamma^\ast}{\Gamma} = \left( 1 + z - \frac{z^2}{4} + \frac{z^3}{8} -
                              \frac{5}{64} z^4 + \ldots \right) .
 \label{31}
\end{equation}
It is therefore clear that the conditional average density $\Gamma^\ast$
and the conditional density $\Gamma$ are equal to each other at the
zeroth order of approximation in the Einstein-de Sitter model, that is,
at the Newtonian approximation. Hence, in our local region the two
functions will basically give the same information about the distribution
of density, and will start to differ outside our local homogeneous region.
However, the ratio $\Gamma^\ast / \Gamma$ does not provide a good
quantitative indicator for the end of our local region since both
$\Gamma^\ast$ and $\Gamma$ will change outside our neighbourhood, with
their ratio being unable to be an accurate gauge of where such a departure
from our local apparent homogeneity will occur.

\section{Conclusion and Discussion}

In this paper I have studied along the past light cone the two-point
correlation function, and a different statistical analysis proposed
by Pietronero (1987), for the dust distribution in the unperturbed
Einstein-de Sitter cosmological model. By using observational relations
already derived in Ribeiro (1992), I explicitly obtained the forms for
the correlation function, the conditional density, and the conditional
average density in the backward null cone, where they are astronomically
observable. By comparing the average number density at our local
time hypersurface and its change along the past null geodesic, it was
possible to obtain explicit quantitative limits for the observable
local homogeneity in the model, and the results showed that the possible
detection of apparent homogeneity are at lower redshift values than usually
assumed, being at $z \approx 0.1$ if a 25\% - 30\% error margin in the
measurements of this density are taken into consideration. Such a
departure from apparent homogeneity occurs well within the region of validity
of the Hubble law, which is a first order linear approximation for the
redshift-distance relation in this cosmology. This happens because the
approximation of local homogeneity in this model is a zeroth
order approximation, while the Hubble law is a first order one.

By studying the behaviour of the correlation function along the past
null geodesic, the results indicate that it does not seem able to fully
characterise the density distribution of the model, being in fact a poor
statistical probe for the behaviour of the average density of the model,
while Pietronero's
(1987) conditional average density seems to fulfill this task as a
statistical test which can be directly derived from galaxy catalogues.
On the other hand, Pietronero-Wertz conditional density seems unable
to characterise quantitatively the departure from apparent homogeneity of the
model, although it will give a qualitative indication of such an effect.
Both functions, however, will basically give the same statistical
information within our local region.

Finally, these results were obtained by using the method of
expanding the relativistic observational relations in power series,
a method in the spirit of the pioneering work of Kristian \& Sachs (1966).

The results of this paper can be interpreted in a variety of ways. First
of all, they do not disprove the Friedmann models in the sense that we
may still be observing galaxies in our present time hypersurface, and so
the homogeneous region could still be within reach.\footnote{ \ Strictly
speaking, we never observe galaxies in our present time hypersurface.
Consequently, what we really do is an ``effective observation'' at our
hypersurface, which is always limited to the error margins. If observations
were perfect, without errors, we would never be able to do calculations
and data reduction as if there were no past light cone problem, but since
observations do have error margins we can approximate and consider an
effective observation at the present time hypersurface provided we always
keep in mind that this is an approximation valid only within the error
margins considered.} However, this possibility
can only be realised within the Friedmann models if either $H_0$ or
$\Omega_0$ or both are smaller than the values used in this paper. For
instance, as we know that in open Friedmann models the scales where the
observable homogeneity finishes is deeper than in the Einstein-de Sitter
cosmology (Ribeiro 1993), we can envisage an open model with a value of
$\Omega_0$ such that the observable homogeneous region ends at much deeper
redshift values than $z=0.1$, as opposed to the case of the $\Omega_0 = 1$
model studied in this paper. If $H_0$ is found to be in the lowest
limits of its present uncertainty, the observable homogeneous region
could still be pushed to even further depths. A closed model seems at
first very difficult to be reconciled with the presently available
astronomical data as it would bring such scales to much closer distances,
even closer than the ones shown in figure 1 for the Einstein-de Sitter
model. Therefore, the only Friedmann models that seem able to be reconciled
with current expectations of an yet to be reached homogeneity are the
ones with low $\Omega_0$ and $H_0$, and this would favor open models
in such scenarios.\footnote{ \ To summarize, if $H_0 > 75 \ \mbox{Km} \
{\mbox{s}}^{-1} {\mbox{Mpc}}^{-1}$, the curves of figures 1 and 3 are
contracted. If $H_0 < 75 \ \mbox{Km} \ {\mbox{s}}^{-1} {\mbox{Mpc}}^{-1}$,
the same curves are stretched.}

These results also seem to make a fractal universe a more attractive
possibility (Pietronero 1987; Ruffini, Song \& Taraglio 1988; Coleman
\& Pietronero 1992; Ribeiro 1993, 1994) since at small scales a fractal
description of the observed lumpiness of matter seems to be a quite
good approximation (Pietronero 1987; Peebles 1989), and as at deeper
ranges we start to observe a deviation from apparent homogeneity in
Friedmann models, we can envisage a single fractal or multifractal
pattern going further, up to the present scales of observations, or,
perhaps, even beyond.

On the problem of relativistic effects in observational cosmology,
the current practice is such that it implicitly leads to a situation
where the assumption that relativistic effects can be ignored in the
study of galaxy correlation and redshift surveys remains basically
untested. Nevertheless, from the results
given above, it seems reasonable to say that we cannot
ignore space curvature and expansion in cosmology at close ranges where
it has been usually thought this can be done without penalty (Peebles
1980, p.\ 143). It is therefore clear that relativistic effects start
to play a role at the first order approximation of at least some
observational relations, and we can only disregard such effects at
the zeroth order of approximation. In other words, the Hubble law is
basically a first order approximation while an Euclidean treatment
of cosmology is good enough only up to zeroth order of
approximation.~\footnote{ \ By Euclidean treatment of cosmology I mean
the use of common Euclidean geometry and the assumption that the dynamical
theory for light given by the theory of relativity can be ignored.}
This is an important point since the usual thinking implicitly 
assumes that the first order approximation is well enough for an
Euclidean treatment in cosmology, and the analysis of this paper shows
that such an assumption may lead to wrong results and misleading
interpretation because current reduction and analysis of data obtained
from not so deep surveys may be subject to the real possibility of
systematic errors. Therefore, the statement ``at small redshifts'',
widely encountered in the literature as implicitly meaning that
``relativistic effects can be ignored at these scales'', should be
used with much greater care.

Finally, from the point of view of a statistical analysis of data,
since figures 1 and 3 show the theoretical form of the number density
distribution in the Einstein-de Sitter model, by accessing the
astronomical data we could in principle compare the theoretical
predictions with observations in order to see if this suffice to
prove or disprove the hypothesis of spatial homogeneity of this
model. This way of looking at the results of figures 1 and 3 has
the advantage of making the concept of spatial homogeneity testable as
$\Gamma$ and $\Gamma^\ast$ are better descriptors of the average density
than $\xi$, and could provide a sort of theoretical fingerprint for
the Einstein-de Sitter model which can be tested observationally.

  \begin{flushleft}
  {\large \bf Acknowledgements}
  \end{flushleft}
  \vspace{5mm}

  ~~~~~Special thanks go to M.\ A.\ H.\ MacCallum for the initial
  discussions which motivated this paper, for the critical reading of
  the manuscript, and for useful suggestions. I also thank R.\ T.\ Jantzen
  for frank discussions about distances in cosmology, S.\ F.\ Rutz for a
  suggestion concerning the algebra of the problem, and the referee for
  suggestions which improved the paper. The financial support from CNPq is
  also acknowledged.

\vspace{20mm}
\begin{flushleft}
{\large \bf References}
\end{flushleft}
  \begin{description}
    \item  Coleman, P. H., \& Pietronero, L. 1992, Phys. Reports, 213, 311
    \item  Coleman, P. H., Pietronero, L., \& Sanders, R. H. 1988, Astron.
           Astrophys., 200, L32
    \item  Davis, M., et al. 1988, Ap. J., 333, L9
    \item  Einasto, J., Klypin, A. A., \& Saar, E. 1986, M. N. R. A. S.,
           219, 457
    \item  Ellis, G. F. R. 1971, in Relativistic Cosmology, Proc.\ of
	        the International School of Physics ``Enrico Fermi'', General
                Relativity and Cosmology, ed.\ R.\ K.\ Sachs (New York: Academic
	        Press), p.\ 104
    \item  Ellis, G. F. R. 1987, in Theory and Observational Limits in
                Cosmology, Proc.\ of the Vatican Observatory Conference held in
                Castel Gandolfo, ed.\ W.\ R.\ Stoeger (Specola Vaticana), p.\ 43
    \item  Geller, M. 1989, in Astronomy, Cosmology and Fundamental
           Physics, ed. M.\ Caffo, R.\ Fanti, G.\ Giacomelli \& A.\ Renzini
           (Dordrecht: Kluwer Academic Publishers), p.\ 83
    \item  Kristian, J., \& Sachs, R. K. 1966, Ap. J., 143, 379
    \item  MacCallum, M. A. H. 1987, in Theory and Observational Limits in
                Cosmology, Proc.\ of the Vatican Observatory Conference held in
                Castel Gandolfo, ed.\ W.\ R.\ Stoeger (Specola Vaticana),
		p.\ 121
    \item  Mart\'{\i}nez, V. J., Portilla, M., Jones, B. J. T., \& Paredes,
                S. 1993, Astron. Astrophys., 280, 5
    \item  McVittie, G. C. 1974, Q.\ Journal R.\ A.\ S., 15, 246
    \item  Peebles, P. J. E. 1980, The Large-Scale Structure of the
           Universe (Princeton: Princenton University Press)
    \item  Peebles, P. J. E. 1989, Physica D, 38, 273
    \item  Pietronero, L. 1987, Physica A, 144, 257
    \item  Ribeiro, M. B. 1992, Ap.\ J., 395, 29 
    \item  Ribeiro, M. B. 1993, Ap.\ J., 415, 469
    \item  Ribeiro, M. B. 1994, in Deterministic Chaos in General
           Relativity, ed.\ D.\ W.\ Hobill, A.\ Burd \& A.\ Coley
	   (New York: Plenum Press), p.\ 269
    \item  Ruffini, R., Song, D. J., \& Taraglio, S. 1988, Astron.
           Astrophys., 190, 1
    \item  Saunders, W., et al. 1990, M. N. R. A. S., 242, 318
    \item  Wertz, J. R. 1970, PhD thesis, University of Texas at Austin
    \item  Wertz, J. R. 1971, Ap.\ J.\, 164, 277
  \end{description}
\newpage
\begin{center}
 {\large Table Captions}
 \vspace{10mm}
\end{center}
\begin{description}
  \item[{\rm Table 1:}] Simple numerical estimates showing that the
                        deviations from the linearity of the
			redshift-distance relation seem to be a
			second order effect, while the departures
			from the local spatial homogeneity are
			a first order effect. The figures show that in
			the Einstein-de Sitter model with
			$H_0=75 \ \mbox{Km} \ {\mbox{s}}^{-1} 
			{\mbox{Mpc}}^{-1}$, relativistic effects would
			start to matter at $z \approx 0.04$, which is
			also the range where the zeroth order
			locally homogeneous approximation would start to be no
			longer valid.

  \item[{\rm Table 2:}] Comparison between the percentage drop from the
                        100\% value at $t=0$, for $\langle n \rangle$
			and $\Gamma$. Considering that the average
			number density gives the actual distribution of
			dust of the Einstein-de Sitter model, the
			conditional density (third column) does not
			seem to be much sensitive to the drop of the
			density itself, although it does indicate such
			decrease qualitatively.
\end{description}

\vspace{1cm}

\begin{center}
 {\large Figure Captions}
 \vspace{10mm}
\end{center}
\begin{description}
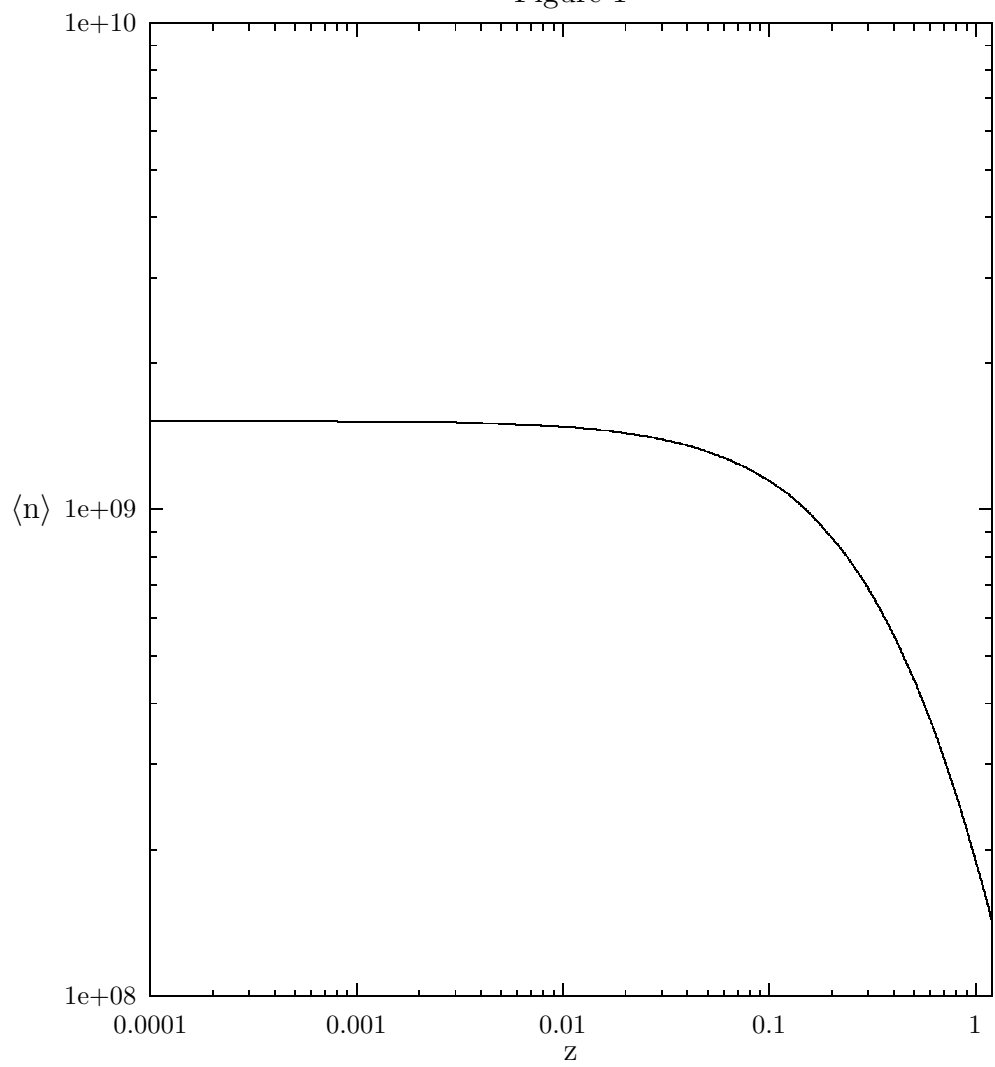
 \item[{\rm Fig. 1:}] Plot of average number density $\langle n \rangle$ vs.
                      the redshift $z$ where the value $H_0=75 \ \mbox{Km} \
		      {\mbox{s}}^{-1} {\mbox{Mpc}}^{-1}$ for the Hubble
                      constant was
		      assumed. This figure is similar to the
                      one presented in Ribeiro (1992) and shows the
                      drop in the density with the redshift in the
                      Einstein-de Sitter model. Note that this drop
                      becomes significant even when $z \ll 1$, although
		      at this range the redshift-distance law is well
		      approximated by a linear function.

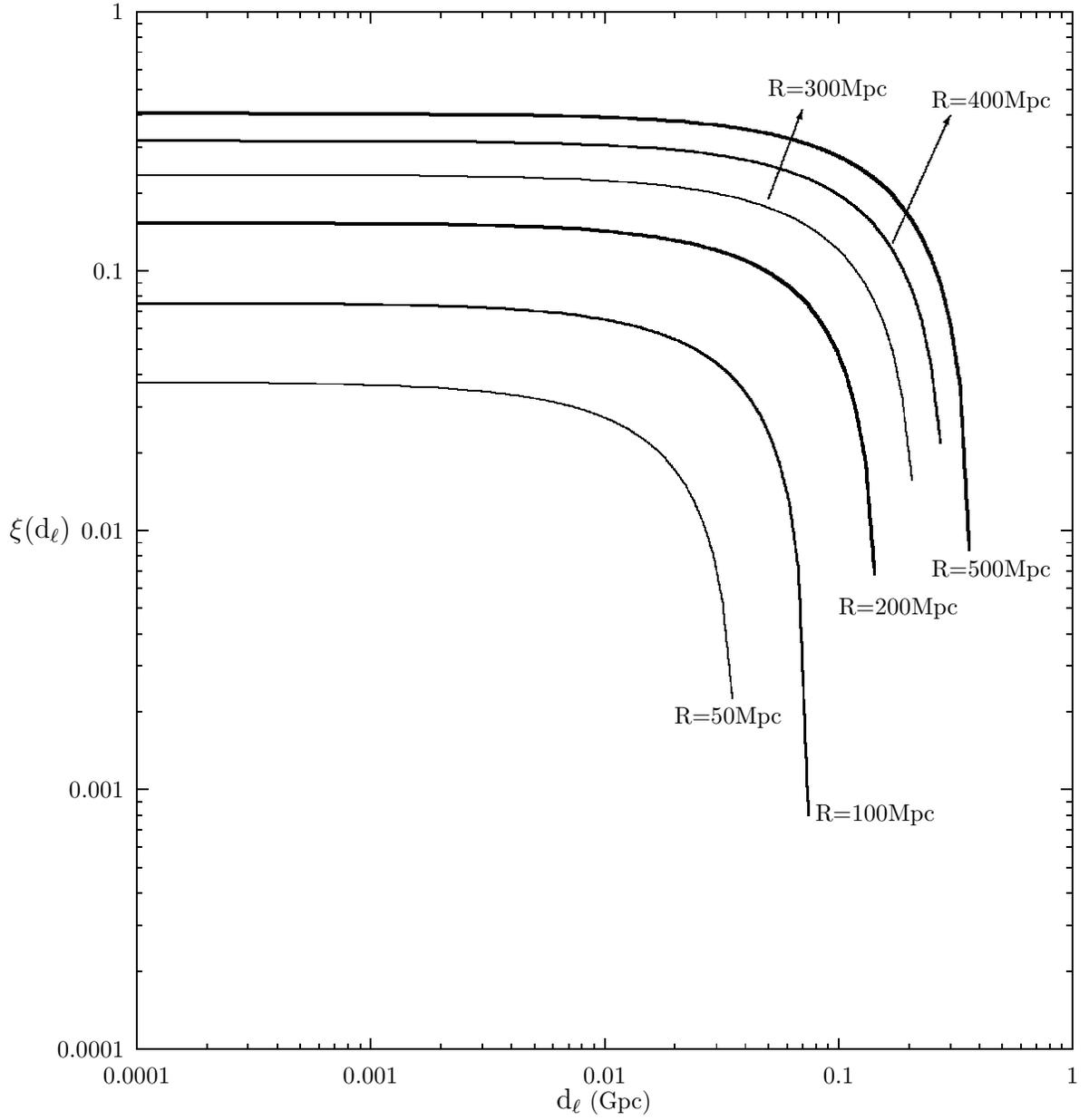
 \item[{\rm Fig. 2:}] The correlation function $\xi (d_\ell)$ plotted for
                      different sample sizes $R$. The graph indicates that
 		      for each different value of $R$ the correlation
		      function behaves as if it had obtained a different
		      scale length for the breakdown of the observable
		      local homogeneity, in contrast to the values of
		      table 1 for the real distribution of dust.

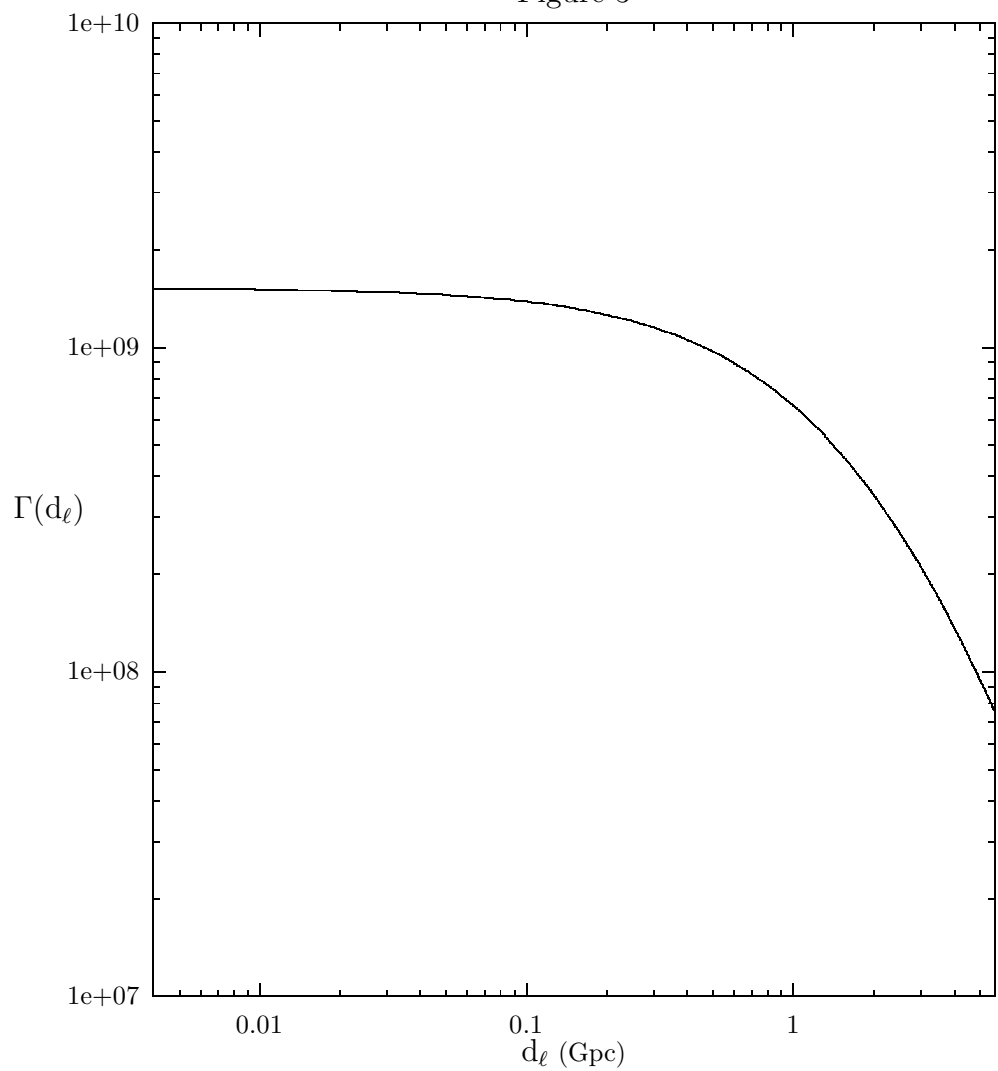
 \item[{\rm Fig. 3:}] The conditional density $\Gamma (d_\ell)$ for the
		      distribution of dust in the Einstein-de Sitter model
		      (eq.\ [\ref{15}]), plotted in the same range as in
		      the case of the average number density of figure 1.
		      One can see that this function traces qualitatively
		      the tendency of the dust of deviating from local
		      homogeneity as the distance increases (shown in
		      figure 1).

\end{description}
\newpage
~\hspace{3cm}\raisebox{-10cm}{
\begin{minipage}{7.4cm}
  {\Large
  \begin{center} Table 1 \end{center}
  \begin{center}

\end{center}
\end{figure}

\end{document}